\def\year{2022}\relax
\documentclass[letterpaper]{article} % DO NOT CHANGE THIS
\usepackage{aaai22}  % DO NOT CHANGE THIS
\usepackage{times}  % DO NOT CHANGE THIS
\usepackage{helvet}  % DO NOT CHANGE THIS
\usepackage{courier}  % DO NOT CHANGE THIS
\usepackage[hyphens]{url}  % DO NOT CHANGE THIS
\usepackage{graphicx} % DO NOT CHANGE ETHIS
\usepackage{natbib}  % DO NOT CHANGE THIS AND DO NOT ADD ANY OPTIONS TO IT
\usepackage{caption} % DO NOT CHANGE THIS AND DO NOT ADD ANY OPTIONS TO IT
\DeclareCaptionStyle{ruled}{labelfont=normalfont,labelsep=colon,strut=off} % DO NOT CHANGE THIS
\usepackage{bbding}
\usepackage{booktabs}
\usepackage{multirow}
\usepackage{amsmath}
\usepackage{algorithm}
\usepackage{algorithmic}
\usepackage{url}
\usepackage[colorinlistoftodos,prependcaption,textsize=tiny]{todonotes}

\usepackage[colorinlistoftodos,prependcaption,textsize=tiny]{todonotes}

\usepackage{newfloat}
\usepackage{listings}
\floatstyle{ruled}
\newfloat{listing}{tb}{lst}{}
\floatname{listing}{Listing}
\nocopyright
\title{HelixFold: An Efficient Implementation of AlphaFold2 using PaddlePaddle}
\author{
    Guoxia Wang, Xiaomin Fang, Zhihua Wu\\
    Yiqun Liu, Yang Xue, Yingfei Xiang\\
    Dianhai Yu, Fan Wang, Yanjun Ma
}
\affiliations{
    Baidu Inc.
}

\usepackage{bibentry}
\begin{document}

\maketitle

\begin{abstract}
Accurate protein structure prediction can significantly accelerate the development of life science. The accuracy of AlphaFold2, a frontier end-to-end structure prediction system, is already close to that of the experimental determination techniques. Due to the complex model architecture and large memory consumption, it requires lots of computational resources and time to implement the training and inference of AlphaFold2 from scratch. The cost of running the original AlphaFold2 is expensive for most individuals and institutions. Therefore, reducing this cost could accelerate the development of life science. We implement AlphaFold2 using PaddlePaddle, namely HelixFold, to improve training and inference speed and reduce memory consumption. The performance is improved by operator fusion, tensor fusion, and hybrid parallelism computation, while the memory is optimized through Recompute, BFloat16, and memory read/write in-place. Compared with the original AlphaFold2 (implemented with Jax) and OpenFold (implemented with PyTorch), HelixFold needs only 7.5 days to complete the full end-to-end training and only 5.3 days when using hybrid parallelism, while both AlphaFold2 and OpenFold take about 11 days. HelixFold saves 1x training time. We verified that HelixFold's accuracy could be on par with AlphaFold2 on the CASP14 and CAMEO datasets. HelixFold's code is available on GitHub for free download: \url{https://github.com/PaddlePaddle/PaddleHelix/tree/dev/apps/protein_folding/helixfold}, and we also provide stable web services on \url{https://paddlehelix.baidu.com/app/drug/protein/forecast}.
\end{abstract}

\section{Introduction}
Proteins are exceptionally critical for life science, as it plays a wide range of functions in organisms. A protein comprises a chain of amino acid residues and folds into a 3D structure to play its functions. Since the 3D structure determines the protein's functions, studying the 3D structure is helpful in understanding the mechanism of the protein's activities. However, it is time-consuming and complex to study protein structure determination through experimental technologies, e.g., X-ray crystallography and nuclear magnetic resonance (NMR). The experimental technologies have determined about one hundred thousand protein structures \cite{sussman1998protein}, but the structures of hundreds of millions of proteins with determined amino acid sequences are still waiting to be determined. Therefore, efficient protein structure estimation methods are in great demand.

Many institutions \cite{jumper2021highly,yang2015tasser,du2021trrosetta,baek2021accurate,peng2011raptorx} made their efforts to develop AI-based protein structure prediction systems due to the efficiency and the capacity of the deep neural networks. In particular, thanks to the fantastic performance in the challenging 14th Critical Assessment of protein Structure Prediction (CASP14 \cite{moult2005decade}), AlphaFold2 \cite{jumper2021highly} from DeepMind has attracted lots of public attention. The accuracy of AlphaFold2 approaches that of the experimental determination technologies. AlphaFold2 is an end-to-end protein estimation pipeline that directly estimates the 3D coordinates of all the atoms in the proteins. A novel and well-designed architecture is proposed to promote the estimation accuracy, which jointly models multiple sequence alignments (MSAs) for evolutional relationships and pairwise relations between the amino acids to learn the spatial relations. 

Although the accuracy of AlphaFold2 is satisfactory for protein structure prediction, it is still challenging for most individuals and institutions to apply AlphaFold2 to study proteins in practice. First, only the inference pipeline of AlphaFold2 is open-sourced. For those who need to train their versions of AlphaFold2, several modules remain to be implemented. A few teams, including OpenFold \cite{Ahdritz_OpenFold_2021}, UniFold, MindSPONGE, and FastFold \cite{cheng2022fastfold}, have attempted to reproduce the complete training pipeline of AlphaFold2 through various deep learning platforms. However, due to the complex model architecture and large memory consumption, training AlphaFold2 from scratch and inferring the structures of the protein with long chains requires lots of computation (e.g., training on 128 TPUv3 cores for 11 days). Apart from computational expenses, it is also challenging to achieve comparable accuracy with AlphaFold2, limiting the its wide usage. Therefore, a computationally efficient implementation of AlphaFold2 with high accuracy would greatly benefit individuals and institutions involved in protein studies.

To this end, we develop HelixFold, an efficient protein structure prediction system implementing AlphaFold2 with PaddlePaddle. HelixFold aims to optimize the computation and memory consumption required for AlphaFold2, providing the complete pipeline of training and inference. We reduce the computation cost by operator fusion, tensor fusion, and hybrid parallelism computation. For example, 14 small operators of complex Gated Self-Attention in Evoformer (a module to exchange the information between the residues in AlphaFold2) are fused into a large C++ operator. 4630 small tensors are fused into only a few large tensors. Furthermore, we propose a new parallelism technique called Branch Parallelism (BP) for distributed parallel computing. BP distributes the computing branches (MSAStack branch and PairStack branch) to different devices to drive up the computing speed. Meanwhile, BP can be combined with other parallel strategies (e.g. data parallelism, DP), and we attempt to apply the hybrid parallelism to train AlphaFold2. For memory optimization, we use multiple common techniques, such as Recompute, BFloat16, and memory read/write in-place.

We compare HelixFold with several versions of AlphaFold2, including the original AlphaFold2 implemented with Jax reported in the paper and OpenFold implemented with PyTorch. For the computational cost, HelixFold takes only 7.5 days to complete the entire end-to-end training process, containing the initial training and fine-tuning stages. When using hybrid parallelism, time consumption can be further reduced to 5.3 days. In comparison, the original AlphaFold2 and OpenFold need about 11 days. We also compare the performance brought by fusion technology and parallel computing for AlphaFold2 and the parallel efficiency of various parallelisms, such as DP, dynamic axial parallelism (DAP), BP, and hybrid parallelism. In the initial training stage, the parallel efficiencies of DP, BP, and DAP reached 94.77\%, 64.43\%, and 43.06\% respectively.
%In the fine-tuning stage, the parallel efficiency of DP is as high as 98.85\%, while the hybrid parallelism of BP and DAP is only 46.41\%.
With a total improvement of 124.48\%, tensor fusion and operator fusion see an improvement in performance by 47.16\% and 42.66\%, respectively. Besides, to verify the accuracy of HelixFold, we trained HelixFold from scratch. The TM-score of converged HelixFold is 87.7 on CASP14, including 87 proteins, and 88.8 on CAMEO, including 371 collected proteins. The experimental results show that HelixFold’s accuracy can be on par with the original AlphaFold2.

The main contributions of HelixFold can be summarized as follows:
% \begin{itemize}
%     \item Optimizing the performance of AlphaFold2 significantly through fusion techniques, including operator fusion and tensor fusion;
%     \item Further improve the performance with hybrid parallel distributed computing. We propose branch parallelism for the branch structure of the Evoformer module, which has high parallel efficiency and low communication overhead;
%     \item Optimizing common memory to reduce memory peaks and run successfully fine-tuning on A100 40G, including Recompute technology, BFloat16, Subbatch, and memory read/write in-place;
%     \item We trained HelixFold from scratch, and the experimental results show that HelixFold’s accuracy can be on par with the original AlphaFold2.
% \end{itemize}
\begin{itemize}
    \item We implemented and released the complete training and inference pipelines of AlpahFold2 with several improvements using PaddlePaddle and name it HelixFold. A stable protein structure prediction service is also available on the website PaddleHelix.
    \item We proposed a novel Branch Parallelism, and improved the computational performance as well as optimized the memory consumption by multiple advanced high-performance computing techniques, requiring only five days to complete the training of AlphaFold2.
    \item We trained HelixFold from scratch, and the experimental results demonstrate that HelixFold’s accuracy can be on par with the original AlphaFold2.
\end{itemize}

\begin{table}[t]
\centering
\resizebox{.95\columnwidth}{!}{
\begin{tabular}{cccc}
   \toprule
   & AlphaFold2 & ViT-B/16 & GPT-3/1.3B \\
   \midrule
   Sequence Shape & (5120, 384), (384, 384) & 192 & 1024 \\
   Hidden Dim & 64 or 128 or 256 & 768 & 2048 \\
   Batch Size per Device & 1 & 128 & 2 \\
   Attention with Gating & \Checkmark & \XSolidBrush & \XSolidBrush \\
   Num of Head & 4 or 8 & 12 & 16 \\
   Blocks & 68=8+4+48+8 & 12 & 24 \\
   Num of Recycle & 4 & - & - \\
   Num of Parameter Elements & 93M & 87.16M & 1.3B \\
   Num of Parameter Tensors & 4630 & 154 & 292 \\
   Memory Peak in Block & 22.5 GiB & 3.375 GiB & 0.125 GiB \\
   Operator Calls per Step & 214606 & 2210 & 3241 \\
   \bottomrule
\end{tabular}
}
\caption{The configuration difference between AlphaFold2 and ViT-B/16 and GPT-3/1.3B. AlphaFold2 has a larger memory overhead and CPU scheduling overhead than the others.}
\label{tab_diff}
\end{table}

\section{Challenges}
\begin{figure*}[t]
\centering
\includegraphics[width=0.98\textwidth]{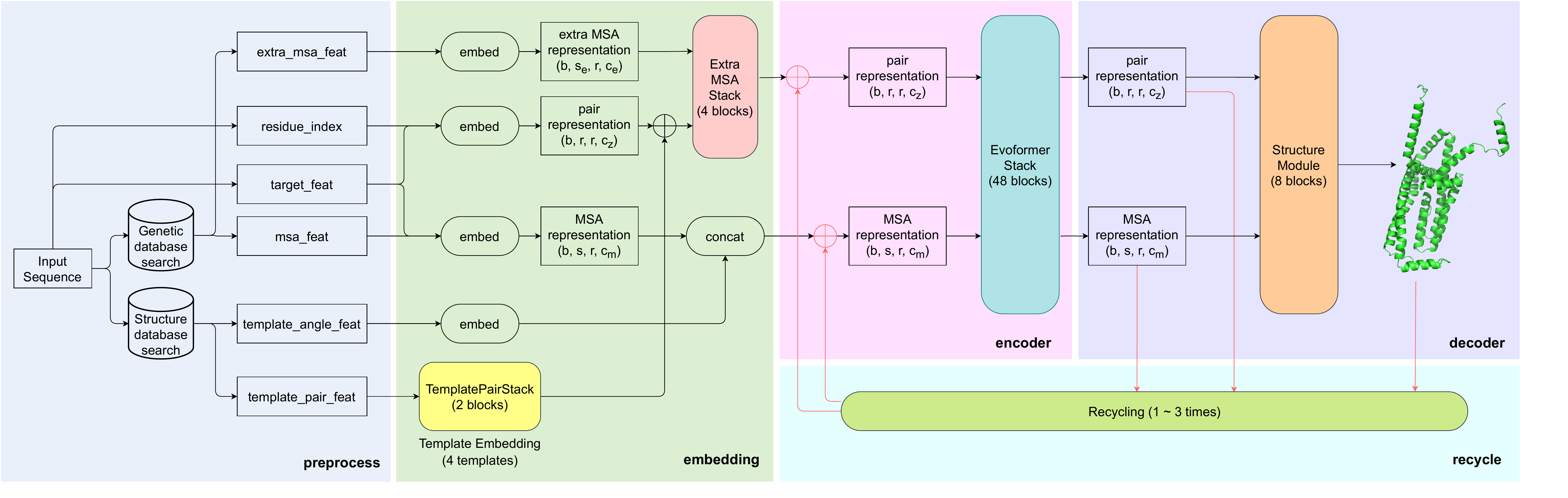} 
\caption{Overall framework of AlphaFold2. Dimension names: $b$: mini-batchs, $s$: clustered MSA sequences, $s_e$: extra MSA sequences, $r$: residues, $c$: channels.}
\label{fig:alphafold2}
\end{figure*}

AlphaFold2 proposed a novel model architecture and training procedure for protein folding. As shown in Figure~\ref{fig:alphafold2}, the model architecture of AlphaFold2 is ingenious but also complicated, with 68 blocks, including 8 \emph{TemplatePairStack} blocks, 4 \emph{ExtraMSAStack} blocks, 48 \emph{EvoformerStack} blocks, and 8 \emph{StructureModule} blocks. Besides, AlphaFold2 executes the forward network multiple times by recycling technique to make the network deeper, where the number of blocks increases from 68 to 236 when $N_{cycle}=4$. By comparing the configurations of AlphaFold2 and two widely used Transformer-based models, i.e., ViT-B/16 and GPT-3/1.3B, as shown in Table~\ref{tab_diff}, we can observe that the high memory consumption and operator scheduling cost are the main challenges of the implementation of AlphaFold2.

\begin{itemize}
\item \textbf{High Memory Consumption} Alphafold2 adopts two tracks to model the proteins: the MSA track and the pairing track. For the MSA track, compared with commonly used models that only store the latent representations of a single sequence, the MSA track in AlphaFold2 stores the latent representations of 512 sequences. For the pairing track, the intermediate memory required by a general model is proportional to the length of the sequence, while the memory required by the pairing track in AlphaFold2 is proportional to the square of the sequence length. Additionally, AlphaFold2 not only contains more than two hundred blocks that are much deeper than the mainstream models but also accepts 4-dimensions input that requires more memory consumption. Consequently, optimizing the memory consumption to feed into the memory of the GPU cards is a great challenge.

\item \textbf{High Operator Scheduling Cost} The model architecture of AlphaFold2 is complicated, and a massive number of calculations are required. Specifically, according to statistics, each step of training needs to schedule hundreds of thousands of operators, which is far more than the tens of thousands of calls of ViT and GPT. In addition, AlphaFold2 contains 4630 parameter tensors, and the cost of traversing and accessing those tensors is considerable. For a deep learning framework based on the dynamic graph, massive operator scheduling brings significant waste to the overall computing performance.

\item \textbf{Extremely Small Batch Size} The total batch size during training is 128 and the batch size is 1 per device. The limited total batch size cannot be scaled to more devices by data parallelism to speed up training, and the batch size on each device cannot be further split.
\end{itemize}

\section{Implementation}

We comprehensively optimized the memory and performance for the implementation of AlphaFold2 based on PaddlePaddle \cite{ma2019paddlepaddle}. We reduce the memory consumption through several commonly used techniques, including Recompute, BFloat16, and Subbatch. Besides, fusion technologies that combine multiple operators and tensors are adopted to alleviate the CPU scheduling cost. Furthermore, we apply hybrid parallelism, which greatly promotes the training throughput of proteins. Especially, we propose a novel parallelism, namely Branch Parallelism, parallelly training the MSA track and Pair track of AlphaFold2 in multiple GPUs.

\subsection{Fused Gated Self-Attention}

\begin{figure}[t]
\centering
\includegraphics[width=0.9\columnwidth]{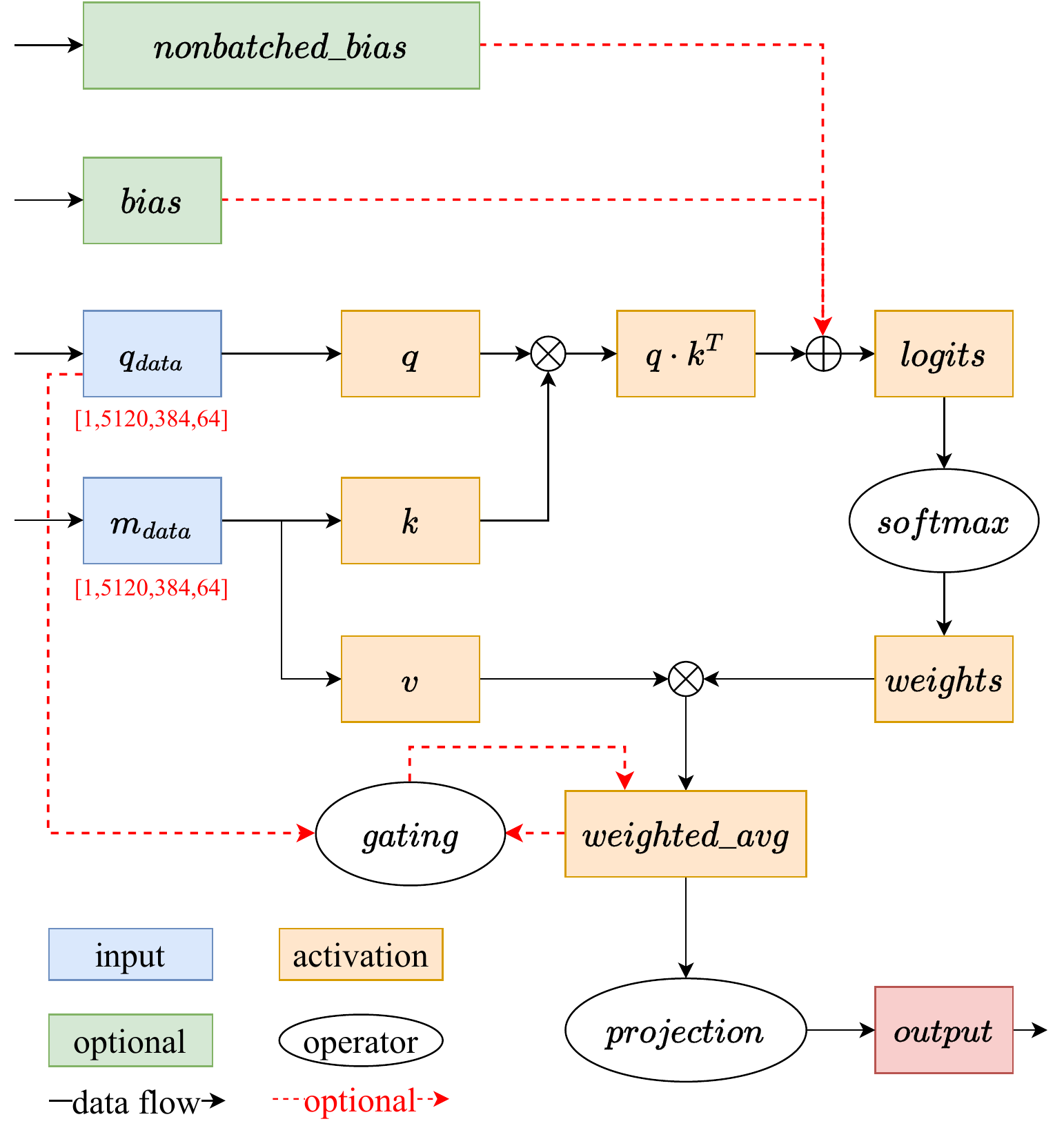} % Reduce the figure size so that it is slightly narrower than the column.
\caption{Gated Self-Attention with pair bias of Evoformer. The 4-dimensional inputs, adding $nonbatched\_bias$ to the logits, and gating mechanism are the main differences from vanilla multi-head attention.}
\label{fig_gate_attention}
\end{figure}

Many Transformer-based systems \cite{rasley2020deepspeed,wang2021lightseq2,fang2021turbotransformers} attempted to fuse the performance-critical multi-head attention (MHA) module into a single operator to increase the throughput.
The attention techniques used in Evoformer of AlphaFold2 are meticulously designed, including row-wise attention, column-wise attention, an optional gating mechanism before the attention output, and additional bias term from pair representation, as shown in Figure~\ref{fig_gate_attention}. Because of the fairly small mini-batch size and sequence length, scheduling a huge number of operators is one of the bottlenecks for training AlphaFold2. The overhead during eager execution mainly comes from (1) the tracing, creating, and destroying of operators for forward and backward propagation, (2) the preparation of input/output tensors, and (3) launches of GPU kernels.
Profiling results show that the average GPU utilization is only 66\% for the initial training iteration.

We implement a Fused Gated Self-Attention operator to optimize both the CPU and GPU overhead in the following manner. 
First, the GEMM computations of \emph{queries}, \emph{keys}, and \emph{values} are merged into a big one with transposed weights. Second, multiple consecutive \emph{element-wise/broadcast} computations are fused in a single GPU kernel. Note that the reduction-based computation, \emph{Softmax}, normally has a limit to the degree of parallelism so that it is not fused with other computations to avoid degrading the performance.

Compared to the straightforward implementation composed of a number of fine-grained operators, the fused coarse-grain gated self-attention only holds five outputs: the transposed output of merged QKV GEMM, the direct output of \emph{Softmax}, attention context, gating project, and the final output. These five outputs will be referenced to calculated gradients, temporary tensors are used for other intermediate results, and they will be released as soon as the related kernels are finished. Cooperated with intra-operator Recompute and memory read/write in-place strategy which will be detailed in the following section, the memory consumed is significantly reduced.

\begin{figure}[t]
\centering
\includegraphics[width=0.98\columnwidth]{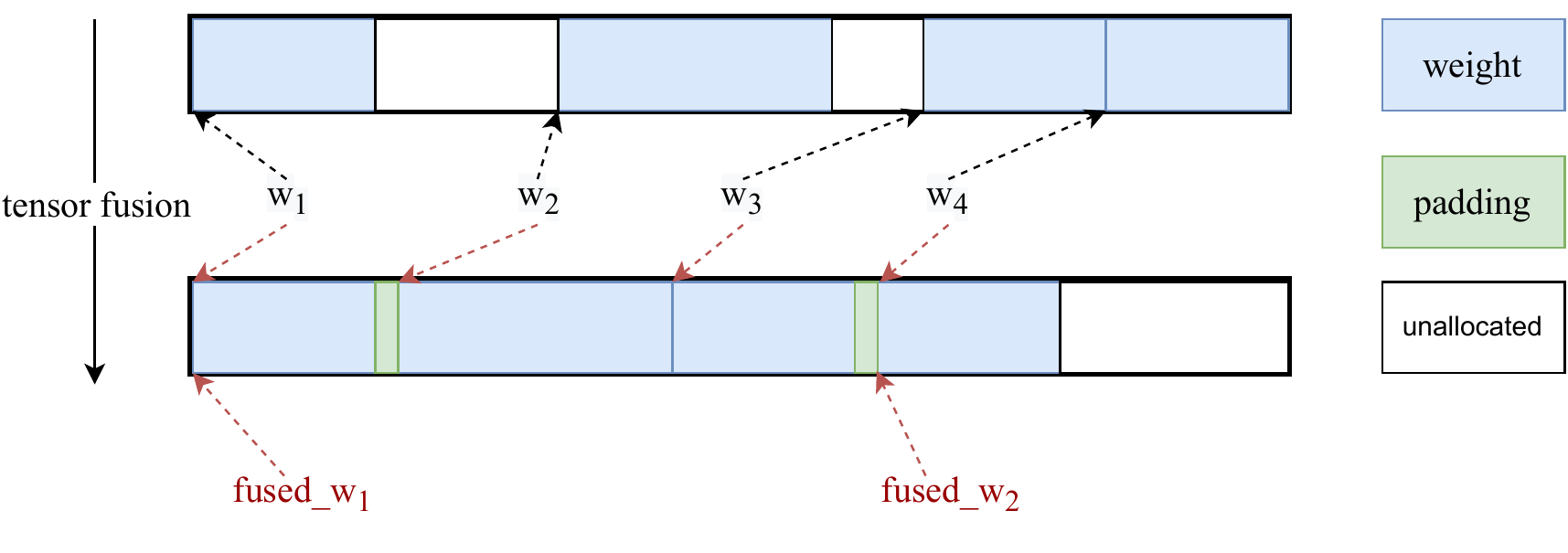} % Reduce the figure size so that it is slightly narrower than the column.
\caption{Tensor Fusion. $w_1$, $w_2$, $w_3$ are fused to $fused\_w_1$ and $w_4$ is fused to $fused\_w_2$. In order to read and write correctly the tensor, we add the padding to align the memory address, e.g. 256 bytes alignment for NVIDIA GPU.}
\label{fig_fusetensor}
\end{figure}

\subsection{Tensor Fusion}

Tensor fusion is a simple and common technique in deep learning. Distributed Data Parallelism fuse multiple gradients to a large tensor and synchronizes them by \emph{Allreduce} operation which is implemented in PyTorch \cite{NEURIPS2019_9015} and PaddlePaddle \cite{ma2019paddlepaddle}. Approaches like DeepSpeed’s ZeRO \cite{rajbhandari2020zero} and FairScale’s Fully Sharded Data Parallel \cite{FairScale2021} fuse not only gradients but also parameters and optimizer states. We use ultimate tensor fusion in the AlphaFold2 to improve performance.

AlphaFold2 has 4630 model parameters. We first fuse these parameters to a single one or a few parameters and modify the data pointer to fused memory. Figure \ref{fig_fusetensor} shows the details of operations. In general, the gradient memory of each model parameter is allocated in the first backward pass. After the parameters are fused, the corresponding fused gradients will be formed based on the fused parameters. Likewise, the gradient pointer of each model parameter will be assigned to the corresponding fused memory address. In Adam optimizer, we do not create optimizer states for each model parameter but fused optimizer states (e.g. momentum and variances). Thus, we fuse parameters, gradients, and optimizer states before the training run.

Table \ref{tab_fusetensor} shows the number of kernel launches, and whether using tensor fusion in several different computing stages, including gradient synchronize, gradient clip by global norm, and update in the optimizer, and parameter update in Exponential Moving Average (EMA). Before using tensor fusion, assuming that the number of parameters of the model is $n$, the number of kernel launches in each computing stage can be expressed as $\mathcal{O}(n)$. As for the gradient clipping stage, it first computes the norm of each gradient and accumulates the global norm, and then clips each gradient. Compared to other stages, each parameter is traversed twice. Therefore, it is inefficient to use separate tensors. The number of kernel launches can reduce to $\mathcal{O}(1)$ when we use tensor fusion. Moreover, tensor fusion can reduce memory fragmentation by reducing the repeated creation and destruction of temporary small tensors.

\begin{table}[t]
\centering
\resizebox{.95\columnwidth}{!}{
\begin{tabular}{ccccc}
   \toprule
   Tensor Fusion & Grad Sync & Grad Clip & Opt Update & EMA \\
   \midrule
   \XSolidBrush & 4630 & 9260 & 4630  & 4630 \\
   \Checkmark & 1 & 2 & 1  & 1 \\
   \bottomrule
\end{tabular}
}
\caption{The number of kernel launches to depend on whether tensor fusion is used.}
\label{tab_fusetensor}
\end{table}

\begin{table}[t]
\centering
\resizebox{.98\columnwidth}{!}{
\begin{tabular}{ccccc}
   \toprule
   Training Process & Tensor Fusion & s/step & protein/s &  \\
   \midrule
   \multirow{2}{*}{Initial training} & \XSolidBrush & 7.79424 & 1.02640 & - \\
   ~ & \Checkmark & 5.96288 & 1.34163 & +30.71\% \\
   \midrule
   \multirow{2}{*}{Fine-tuning} & \XSolidBrush & 18.58428 & 0.43047 & - \\
   ~ & \Checkmark & 16.84065 & 0.47504 & +10.35\% \\
   \bottomrule
\end{tabular}
}
\caption{Speedup using Tensor Fusion for initial training and fine-tuning with data parallelism on 8 GPUs, where batchsize is 1 on each card.}
\label{tab_fusetensor_speed}
\end{table}

\begin{figure*}[t]
\centering
\includegraphics[width=0.96\textwidth]{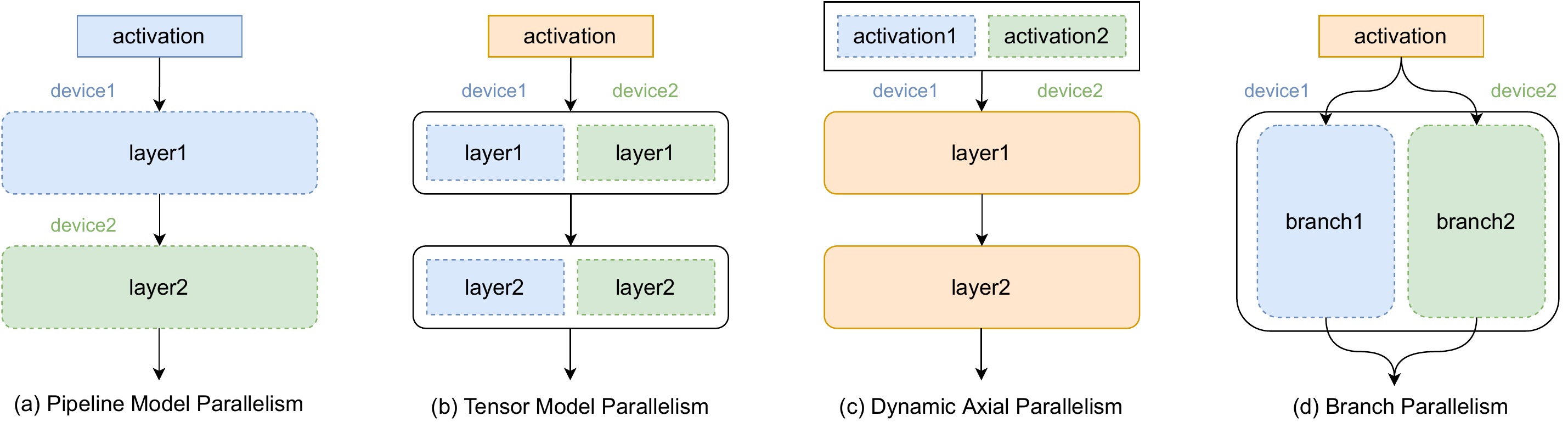} % Reduce the figure size so that it is slightly narrower than the column.
\caption{Various Parallelism Techniques. (a) Tensor Model Parallelism splits the parameter into individual layers to multiple devices at the tensor dimension instead of the input activation; (b) Pipeline Model Parallelism splits the model at the layer dimension; (c) Dynamic Axial Parallelism splits the input activation rather than the model parameter, e.g. the split MSA representation shape is $[B, N_{seq}/N_{device}, N_{res}, C_m]$ in Evoformer; (d) Branch Parallelism splits the calculation branch to multiple devices where each device has the same input activation and parameter.}
\label{fig_parallelisms}
\end{figure*}

\subsection{Hybrid Parallelism Computing}

We first review the various parallelism techniques for training large-scale models in NLP and CV. The mainstream approaches mainly use more accelerators to speed up the training, such as Data Parallelism (DP) \cite{li2020pytorch} and Model Parallelism (MP) \cite{narayanan2021efficient}. Then we introduce the Dynamic Axial Parallelism (DAP) \cite{cheng2022fastfold} proposed for AlphaFold2, which also uses more accelerators to improve training throughput, and we detail its pros and cons. Finally, we propose a novel parallelism technique named Branch Parallelism (BP) to improve training speed for the Evoformer network model structure of AlphaFold2.

\subsubsection{Data Parallelism} is the most popular and efficient method in deep learning distributed training. Each worker has a copy of the model parameters, and parallelism operates in the data dimension. Each worker accepts mini-batch samples and feeds them into the model to calculate the gradient of the parameter through the forward and backward pass. Workers aggregate their gradients by \emph{AllReduce} communication operation to obtain consistent gradients and update parameters. The mini-batch samples are $b$ on each worker and the total mini-batch cross $N$ workers is $B=N*b$. It can speed up training by scaling to more devices at the expense of increasing the total mini-batch samples $B$ in each step.

\subsubsection{Model Parallelism} can be divided into Pipeline Model Parallelism (PP) and Tensor Model Parallelism (TP) in Megatron-LM \cite{narayanan2021efficient}, whose computation and model parameters are distributed to different devices. PP splits the whole model to multiple devices at layer dimension and will introduce the idle time named \textit{pipeline bubble}, Figures \ref{fig_parallelisms}(a). To reduce the size of the pipeline bubble, a mini-batch is split into multiple micro-batches and uses interleaved 1F1B schedule. However, the mini-batch $b=1$ on each worker in AlphaFold2, can not further split into more micro-batches so that the size of the pipeline bubble can not be reduced, resulting inefficient pipeline model parallelism in AlphaFold2. TP distributes the parameter in individual layers to multiple devices at the tensor dimension, Figures \ref{fig_parallelisms}(b). For example, it splits the weight matrix in the feed forward network (FFN) with its column dimension and in the MHA along its head dimension. The whole size of model parameters is only 93M, yet the number of parameters is up to 4630 in AlphaFold2. The hidden size is small (e.g. 64, 128, or 256), while the heads in MHA are 4, which requires significant communication between each layer. As a result, TP has poor computation-communication efficiency.

\subsubsection{Dynamic Axial Parallelism} is proposed to solve the inefficient problem of training the AlphaFold2 model with a small parameter shape and a large activation memory consumption. In particular, Evoformer of AlphaFold2 has two feature inputs which have four dimensions, MSA representation ($[B, N_{seq}, N_{res}, C_m]$) and pair representation ($[B, N_{res}, N_{res}, C_z]$), where $B$ is the batch size, $N_{seq}$ is the length of MSA sequence, $N_{res}$ is the number of residues, $C_m$ and $C_z$ are the channels. DAP splits the activations along the first dimension in both MSA representation and pair representation to multiple devices, Figures \ref{fig_parallelisms}(c). The motivation of DAP is that the calculation mostly occurs in the last two dimensions, e.g. \emph{MSARowAttentionWithPairBias} and \emph{MSAColumnAttention}. It inserts \emph{AlltoAll} and \emph{AllGather} communication to preserve strict calculation semantics. As described in \cite{cheng2022fastfold}, DAP supports the whole Evoformer, while the amount of communication and memory consumption is less than TP, and it can do computation-communication overlap by \emph{Duality Async Operation}. Additionally, it can significantly reduce memory peaks. However, DAP also has several limitations in Evoformer.

\begin{itemize}
    \item The channels are small in MSA and pair representation in Evoformer, and the GPU times are smaller while CPU scheduling overhead is large for the operator, resulting in inefficient performance. Further split activation with DAP would contribute to a very slow speed. The performance will be even lower than without DAP in the initial training stage where $N_{seq}=128$ and $N_{res}=256$;
    \item The communication overhead is large, even though \emph{Duality Async Operation} may cause computation-communication overlap. DAP introduces 6 times \emph{Async AllGather} in the forward and 6 times \emph{Async ReduceScatter} in the backward, 2 times \emph{Async All2All} (forward 1 times, backward 1 times) and 10 times \emph{Sync All2All} (forward 5 times, backward 5 times). The \emph{Async AllGather} and \emph{Async ReduceScatter} can be computation-communication overlap theoretically, but the overlay space is just calculating the $qkv$, $qk^T$ experimentally.
\end{itemize}

\subsubsection{Branch Parallelism}
To this end, we propose Branch Parallelism for the Evoformer of AlphaFold2, Figures \ref{fig_parallelisms}(d). BP is a novel parallelism technique that applies to the AlphaFold2 Evoformer model structure and model structure with multiply branches. The amount of the calculations for each branch is similar. The motivation of BP is that the Evoformer model structure has two calculation branches, namely the MSA stack, and the Pair stack. The devices of the BP group have a copy of parameters and BP splits the calculation branch across multiple devices. One device calculates the MSA stack and the other calculates the Pair stack and inserts \emph{Broadcast} and \emph{AllReduce} communication to preserve strict calculation semantics. 

Specifically, in the forward stage, we do not need to split the input in each Evoformer block as the first device uses the MSA representation to calculate the MSA stack and Outer Product Mean branch then the outputs be broadcast to the second device. The second device uses the pair representation to calculate the Pair stack branch and then broadcast the output to the first device similarly. In the backward stage, we only insert \emph{AllReduce} to accumulate the sum of the gradient of the input pair representation in each Evoformer block. At the end of backward propagation of the whole Evoformer, we need an extra broadcast of the gradient of MSA representation. Like the DAP, we adopt the \emph{AllReduce} or \emph{Broadcast} communication to synchronize the gradients of Evoformer model parameters.

\begin{table*}[t]
\centering
%\resizebox{.95\columnwidth}{!}{
\begin{tabular}{lrr}
   \toprule
    & DAP & BP \\
   \midrule
   MSA stack &  4 $\times$ All2All $+$ 1 $\times$ AllGather $+$ 1 $\times$ ReduceScatter  & 1 $\times$ Broadcast \\
   Pair stack &  8 $\times$ All2All $+$  4 $\times$ AllGather $+$ 4 $\times$ ReduceScatter & 1 $\times$ AllReduce $+$ 1 $\times$ Broadcast \\
   Outer Product Mean & 1 $\times$ AllGather $+$ 1 $\times$ ReduceScatter & 1 $\times$ Broadcast \\
   \midrule
   Total & 24 & 4 \\
   \bottomrule
\end{tabular}
%}
\caption{Communication overhead comparison for each Evoformer module. To simplify the calculation, the actual communication overhead is reduced to the number of invoke of the communication operator.}
\label{tab_communication}
\end{table*}

The detailed communication overhead comparison between DAP and BP is shown in Table \ref{tab_communication}. It can be observed that our proposed BP has only 4 communications per Evoformer block, while the DAP with high communication overhead has 24 complex communications. Note that the communication operator needs extra cost implemented with PyTorch Autogard Function \cite{NEURIPS2019_9015}, e.g. \emph{AllGather} need call \emph{chunk} or \emph{concatenate} function, and we only call \emph{Broadcast} communication operator. For the backward pass of pair representation, we only register a hook function that invokes \emph{AllReduce} communication to synchronize the gradient. 

As discussed above, the Evoformer block comprises many small GPU kernels and CPU launch overheads, which is the performance bottleneck. BP does not split the activation to keep a higher computation-communication efficiency than DAP.

However, BP also has its limitations: it can be scaled up to as many devices as the number of computational branches and requires a similar amount of computation from each branch.

\subsubsection{Hybrid Parallelism} DAP splits the activation to reduce the memory footprint which will be detailed in the memory optimization section. In addition, DAP also has high parallel efficiency with a large shape of input, e.g. in the fine-tuning stage. Therefore, we can combine BP and DAP to train the AlphaFold2 model when CPU launch overheads are not the performance bottleneck. Since BP and DAP just split the computing branches and the activation across multiple devices respectively, the same protein is processed on each device. To improve communication efficiency, BP and DAP work within a single node where the inter-GPU communication bandwidth is high. With data parallelism, we can scale the total mini-batch to 128. We call this technique BP-DAP-DP hybrid parallelism. 

\subsection{Memory Consumption Optimization}

\subsubsection{Recompute} is proposed to train deeper and more complex models in \cite{chen2016training}. It drops some of the intermediate activations and recomputes them by extra forward computation in the backward pass. The memory of AlphaFold2 is mainly consumed on the intermediate activation. In addition, The main trunk of the AlphaFold2 consists of 48 Evoformer blocks. If using the additional model inputs, there are an extra 4 \emph{ExtraMSAStack} blocks and 2 \emph{TemplatePairStack} blocks. In our experiment, we can not run it successfully on NVIDIA A100 (40G) GPU. Shockingly, if we use the Float32 data type, one block per module requires 33.48 GiB.

\begin{table}[t]
\centering
\resizebox{.95\columnwidth}{!}{
\begin{tabular}{cccc}
   \toprule
   Evoformer & ExtraMsaStack & TemplatePairStack & Memory \\
   \midrule
   \XSolidBrush & \XSolidBrush & \XSolidBrush & OOM  \\
   \Checkmark & \XSolidBrush & \Checkmark & OOM  \\
   \Checkmark & \Checkmark & \XSolidBrush & 38.3 GiB  \\
   \Checkmark & \Checkmark & \Checkmark & 19.1 GiB  \\
   \bottomrule
\end{tabular}
}
\caption{Memory consumption whether using Recompute in the initial training stage with Float32.}
\label{tab_recompute}
\end{table}

Like AlphaFold2, we reduce the memory footprint by storing the input activation of each block and recomputing all activations within the blocks in the backward pass. As shown in Table \ref{tab_recompute}, we found that \emph{ExtraMSAStack} and \emph{Evoformer} have too many blocks, and using Recompute on each block can reduce the memory footprint to 38.3 GiB in initial training which can be run on NVIDIA A100 (40G) GPU. Furthermore, we use Recompute on \emph{TemplatePairStack} to reduce the memory consumption to just 19.1 GiB.

\subsubsection {BFloat16} The BFloat16 (Brain Floating Point Format) format was originally developed by Google Brain and available on the second- and third-generation TPU \cite{bf16_tpuv3}. NVIDIA CUDA also supports the BFloat16 format. BFloat16 is a customized 16-bit floating point format and is a truncated IEEE 754 single-precision 32-bit float. It is comprised of 1 sign bit, 8 exponent bits, and 7 significand bits. The dynamic range of BFloat16 is identical to that of Float32. The BFloat16 format improves performance and the memory consumption is reduced by half compared with Float32. It has the same accuracy as mixed-precision training which keeps parameters and gradients in Float32 but converts activations to BFloat16. As we mentioned, AlphaFold2 has large intermediate activations and reduces the memory consumption from 19.1 GiB to 12.7 GiB via BFloat16 for activations in initial training.

In addition, we support BFloat16 format communication in hybrid parallelism. It can reduce the communication volume to improve overall communication speed.

\subsubsection{Memory Read/Write In-place}

\begin{figure}[t]
\centering
\includegraphics[width=0.9\columnwidth]{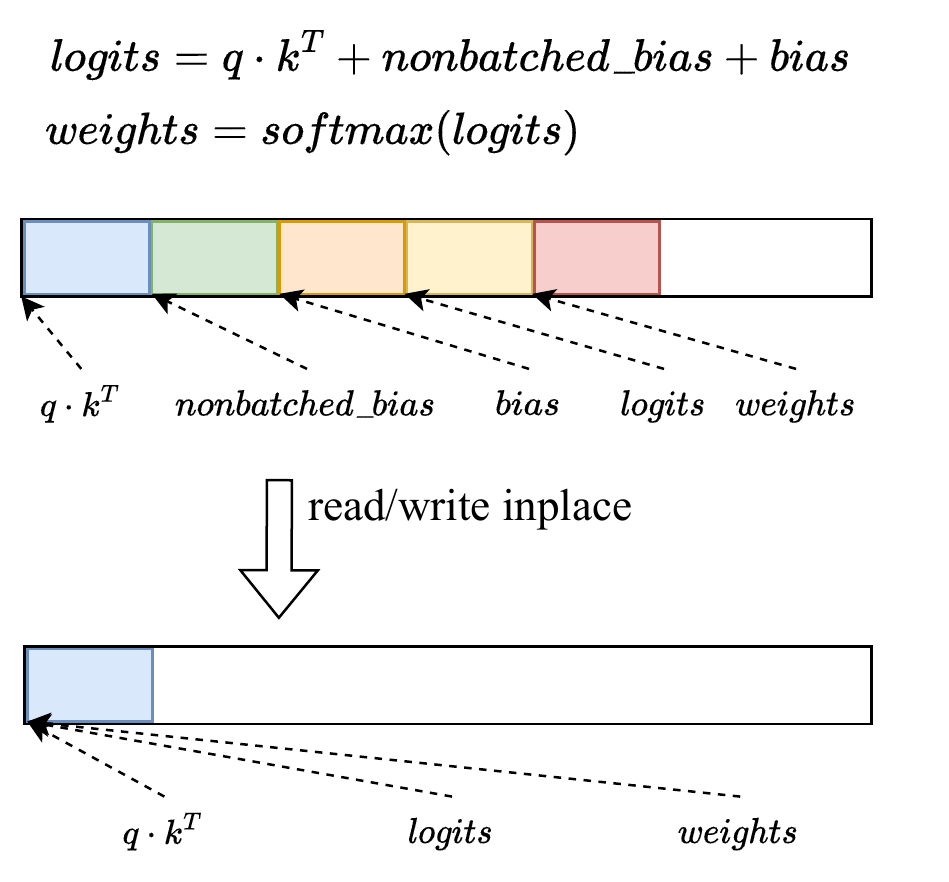} % Reduce the figure size so that it is slightly narrower than the column.
\caption{Memory in-place in fused gate attention. Top: Allocate memory for each intermediate activation; Bottom: Reserve only the memory allocation of $q \cdot k^T$ and read and write in-place on it.}
\label{fig_inplace}
\end{figure}

As we discussed above, we fuse gated self-attention into one C++ Operator. It enables us to optimize the memory consumption of the intermediate activation. As the top of Figure \ref{fig_inplace} shows, in order to calculate the output of \emph{Softmax}, we first need to accumulate $q \cdot k^T$, $nonbatched\_bias$ and $bias$. The memory allocated for each variable will be retained until the gradient is calculated and freed in the backward pass. Note that $nonbatched\_bias$ and $bias$ need to broadcast the shape is the same as the $q \cdot k^T$ and hold the intermediate variable. 

Analysis indicates that it only needs the allocated memory of $q \cdot k^T$, and read/write in-place in the C++ kernel, as shown at the bottom of Figure \ref{fig_inplace}. In addition, we need 2 times memory footprint to calculate the gradient of $logits$ in the backward pass. Therefore, we can reduce memory consumption by 2/3 and reduce memory peaks drastically.

\subsubsection{Subbatch VS DAP} Peak memory consumption is dramatically large in Gated Self-Attention when $N_{seq}$ is very large in fine-tuning or proteins with extremely long amino acid sequences in inference. For example, the input shape of \emph{ExtraMSAStack} in fine-tuning is $[B, N_{seq}, N_{res}, C_m]=[1, 5120, 384, 64]$. Calculating the logits takes the size of
\begin{eqnarray}
\begin{aligned}
logits&=q \cdot k^T  \nonumber \\
&=\underset{N_{seq}}{5120} \cdot \underset{N_{head}}{8} \cdot \underset{N_{res}^2}{384^2} \cdot \underset{sizeof(BFloat16)}{2}\\
&=11.25 \ \text{GiB}
\end{aligned}
\end{eqnarray}
which is too large for training and it will run out of memory in the backward pass. However, the input and output of \emph{ExtraMSAStack} are just 0.2 GiB respectively. We call it inverted hourglass memory where the input and output memory footprint is relatively small while the intermediate activation memory footprint is dramatically large.

\begin{figure*}[t]
\centering
\includegraphics[width=0.96\textwidth]{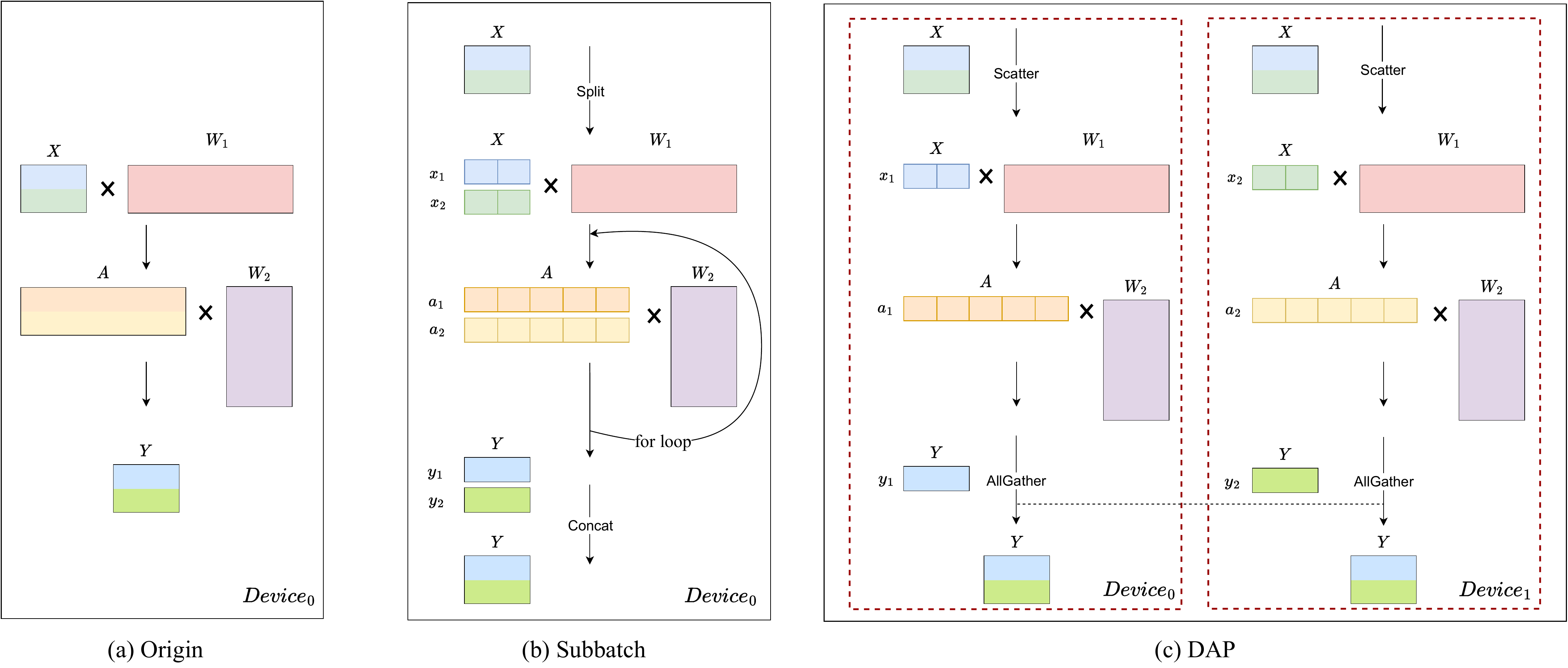} % Reduce the figure size so that it is slightly narrower than the column.
\caption{An simple example of reducing intermediate activation $A$ memory consumption  by Subbatch and DAP. (a) Matmul operator on $A=X*W_1, Y=A*W_2$ with original large activation; (b) Subbatch \emph{Split} the input $X$ to $[x_1, x_2]$ on row dimension and do matmul on sub activation in \emph{for loop} then concatenate the results $[y_1, y_2]$ to the output $Y$; (c) DAP \emph{Scatter} the input $X$ to $[x_1, x_2]$ and do parallel calculation using the corresponding sub activation on each device then obtain the output $Y$ by \emph{AllGather} communication operator.}
\label{fig_subbatch}
\end{figure*}

To solve this problem, Subbatch or Chunking technique in AlphaFold2 and Reformer \cite{kitaev2020reformer} can reduce the peak memory linearly. As Figure \ref{fig_subbatch}(a) shows, a 2 Fully Connected layers example, the input is $X$ with shape $[2,2]$, the output is $Y$ with shape $[2,2]$, $W_1$ and $W_2$ are the parameters of each layer, $A$ is the intermediate activation with shape $[2,5]$. Generally, we call directly $A=X*W_1, Y=A*W_2$, and the peak memory of the intermediate activation is 20 bytes with BFloat16. We identify a ‘batch-like’ dimension of the input and the output for each layer, where the computation is independent in that dimension. Subbatch splits the input $X$ to chunks $[x_1,x_2]$ along row dimension via chunk size equal to 1. Then $x_1$ and $x_2$ are fed into the model respectively in \emph{for loop}. At the end, the results $[y_1,y_2]$ are be concatenated to $Y$. As we can see in Figure \ref{fig_subbatch}(b), the peak memory footprint of the intermediate activation $A$ was reduced to 10 bytes at each iteration in \emph{for loop}. The main notion of Subbatch is to convert the large tensor computation to multiple serial computations and to reduce the peak memory footprint of the intermediate activation.

In parallelism training, we also can distribute each chunk to multiple devices and parallelly computing. The results $y_1$ and $y_2$ are aggregated into the output $Y$ by \emph{AllGather} communication operator. This parallelism technique is called DAP \cite{cheng2022fastfold}, as shown in Figure\ref{fig_subbatch}(c). DAP is another tensor model  parallelism \cite{narayanan2021efficient} that splits the parameter tensor to multiple devices while DAP splits the activation tensor to multiple devices.

In our implementation, we use Subbatch in \emph{MSARowAttentionWithPairBias} for fine-tuning in which the input shape is $[1, 5120, 384, 64]$ and we set the chunk size to 512. We reduce the peak memory footprint from 11.25 GiB to 1.125 GiB. In addition, we use Subbatch and DAP to infer extremely long amino acid sequences of more than 4K in length.

\section{Experiments}

\subsection{Experimental Setup}
\subsubsection{Datasets}
For training, we follow the settings reported in the paper of AlphaFold2 to collect the training data, including 25\% of samples from RCSB PDB (\url{https://www.rcsb.org/}) \cite{10.1093/nar/28.1.235,10.1093/nar/gkaa1038} and 75\% of self-distillation samples. 

For evaluation, we collected two datasets: CASP14 and CAMEO. We collected 87 domain targets from CASP14 (\url{https://predictioncenter.org/casp14/index.cgi}) \cite{jumper2021highly,https://doi.org/10.1002/prot.26202,https://doi.org/10.1002/prot.26237}. We also collected 371 protein targets from CAMEO \url{https://www.cameo3d.org/} \cite{https://doi.org/10.1002/prot.26213}, ranging from 2021-09-04 to 2022-02-19.

\subsubsection{Settings of Model Architectures}
We use two model settings to assess the computational performance of HelixFold and other implementations of AlphaFold2. The model settings is shown in Table \ref{tab:model_settins}, with \emph{initial training} setting corresponding to model 1 and \emph{fine-tuning} setting corresponding to model 1.1.1 reported in the supplementary information of paper AlphaFold2.

\subsubsection{Hyperparameter setting}
All our experiments are run on NVIDIA A100 (40G) and the mini-batch size is 1 on each device. We strictly follow the settings in AlphaFold2. In the initial training stage, we train $10 \times 10^6$ samples (78125 steps), and in the fine-tuning stage, we continue to train $1.5 \times 10^6$ samples (11718 steps). As a feature in AlphaFold2, recycling iterations is to randomize a number from 1 to 4 in each step, perform the forward pass multiple times, and then perform the backward pass. To compare performance quickly and fairly, we firstly fix the random seed, then run 105 training steps, discard the first 5 steps, and finally calculate the average 
speed for the last 100 steps. After our extensive experimental verification, the average speed of 100 steps can get a similar global speed. We fix $random\_seed = 32$, then the random seed for each step is calculated by $random\_seed + cur\_step$. Unless otherwise specified, we use AMP for training, which the parameter using Float32 and intermediate activation using BFloat16.

\begin{table}[t]
\centering
\resizebox{.95\columnwidth}{!}{
\begin{tabular}{cccccc}
   \toprule
   Training Process & Model setting & $N_{templ}$ & $N_{res}$ & $N_{seq}$ & $N_{extra\_seq}$\\
   \midrule
   Initial training & Model 1 & 4 & 256 & 128 & 1024 \\
   Fine-tuning & Model 1.1.1 & 4 & 384 & 512 & 5120 \\
   \bottomrule
\end{tabular}
}
\caption{Model settings of HelixFold for performance comparison.}
\label{tab:model_settins}
\end{table}

\subsection{Accuracy of Prediction}
To verify the effectiveness of HelixFold, we trained it from scratch. The training procedure includes the initial training and the fine-tuning stages. We compare HelixFold and AlphaFold2 (using the publicly available inference pipeline) on datasets CASP14 and CAMEO, as shown in Table~\ref{tab:accuracy}. Four commonly used metrics are used to evaluate the accuracy of HelixFold and AlphaFold2 for protein prediction, including TM-score, LDDT, GDT-TS, and GDT-HA. The results show that HelixFold can achieve competitive accuracy with AlphaFold2.

\begin{table*}[t]
    \centering
    \small
    \begin{tabular}{c|cccc|cccc}
    \toprule
    & \multicolumn{4}{c|}{\textbf{CASP14}} & \multicolumn{4}{c}{\textbf{CAMEO}}\\
    \midrule
    Method & TM-score & LDDT & GDT-TS & GDT-HA & TM-score & LDDT & GDT-TS & GDT-HA\\
    \midrule
    AlphaFold2 & 0.8772 & 0.8021 & 0.8496 & 0.7175 & 0.8862 & 0.8496 & 0.8676 & 0.7546 \\
    HelixFold    & 0.8771 & 0.7932 & 0.8506 & 0.7139 & 0.8885 & 0.8344 & 0.8656 & 0.7436\\
    \bottomrule
    \end{tabular}
    \caption{Comparing accuracy of prediction.}
    \label{tab:accuracy}
\end{table*}

\subsection{Training Performance}

\subsubsection{End-to-end training Training Performance}

HelixFold is efficiently implemented on the PaddlePaddle deep learning framework based on the official open-sourced AlphaFold2 of DeepMind. We made a comparison among AlphaFold2 \cite{jumper2021highly} implemented with Jax and OpenFold \cite{Ahdritz_OpenFold_2021} implemented with PyTorch, in terms of training throughput, total training time, and resource cost, as shown in Table \ref{tab_end2end_performance}. There is a large discrepancy between the input shapes of different training stages, resulting in different throughput and resource costs. Therefore, we firstly compare the performance of each stage in a fine-grained manner.

In the initial training stage, when the
same GPUs (128 $\times$ A100) are used, the throughput of HelixFold is 28.42\% and 52.13\% higher than that of AlphaFold2 and OpenFold respectively. When we use the hybrid parallelism (DP=128, BP=2, DAP=1) on 256 A100 GPUs, HelixFold's throughput is 66.86\% higher than AlphaFold2 and 97.66\% higher than OpenFold. In the fine-tuning stage where the length of the input becomes larger, HelixFold's throughput is 84.49\% higher than AlphaFold2, and 24.03\% higher than OpenFold. When on 512 A100 GPUs using the hybrid parallelism (DP=128, BP=2, DAP=2), HelixFold achieves more than 1x performance improvement, 240.15\% and 128.67\%, respectively.

Compared to the 11 days of total training time needed by AlphaFold2 and OpenFold, HelixFold takes only 7.5 days with the same number of accelerator cards. If with more GPUs to accelerate training, HelixFold only takes 5.3 days, faster than AlphaFold2 by 106.97\% and OpenFold by 104.86\%. It almost cuts the training time by half. 

\begin{table*}[t]
\centering
\resizebox{0.99\textwidth}{!}{
\begin{tabular}{ccccccccc}
   \toprule
   Implementation & Framework & Training Process & Hardward & Step Time (s) & Protein/s & Training Time (days) & Total Training Time (days) & Resource \\
   \midrule
   \multirow{2}{*}{AlphaFold2} & \multirow{2}{*}{JAX} & Initial training & \multirow{2}{*}{128 × TPUv3} & 7.513 & 17.037 & 6.7934 & \multirow{2}{*}{10.961} & \multirow{2}{*}{33672 TPU hours} \\
   ~ & ~ & Fine-tuning & ~ & 30.729 & 4.1654 & 4.1676 & ~ & ~ \\
   \midrule
   \multirow{2}{*}{OpenFold} & \multirow{2}{*}{PyTorch} & Initial training & \multirow{2}{*}{128 × A100(40G)} & 8.9 & 14.382 & 8.0476 & \multirow{2}{*}{10.849} & \multirow{2}{*}{33328 GPU hours} \\
   ~ & ~ & Fine-tuning & ~ & 20.657 & 6.196 & 2.8016 & ~ & ~ \\
   
   \midrule
   \multirow{2}{*}{HelixFold} & \multirow{2}{*}{PaddlePaddle} & Initial training & \multirow{2}{*}{128 × A100(40G)} & 5.849 & 21.880 & 5.2895 & \multirow{2}{*}{7.5483} & \multirow{2}{*}{23188 GPU hours} \\
   ~ & ~ & Fine-tuning & ~ & 16.654 & 7.685 & 2.2588 & ~ & ~ \\
   \midrule
   \multirow{2}{*}{HelixFold$^*$} & \multirow{2}{*}{PaddlePaddle} & Initial training & 256 × A100(40G) & 4.502 & 28.428 & 4.0708 & \multirow{2}{*}{5.2958} & \multirow{2}{*}{40063 GPU hours} \\
   ~ & ~ & Fine-tuning & 512 × A100(40G) & 9.034 & 14.169 & 1.2250 & ~ & ~ \\
   \bottomrule
\end{tabular}
}
\caption{Complete end-to-end training performance and  resource cost comparison. }
\label{tab_end2end_performance}
\end{table*}

\subsubsection{Performance of Fused Gated Self-Attention}
We compare Fused Gated Self-Attention with a direct implementation of Gated Self-Attention consisting of multiple fine-grained operators. As shown in Table \ref{tab_fused_gating_attention_speed}, the performance improvement of Fused Gated Self-Attention is significant in initial training and fine-tuning. Especially in the initial training, the performance is 48.79\% higher than no operator fusion is used, mainly due to the reduced scheduling overhead of the CPU. Several small operators are merged into a C++ operator, and fewer CUDA kernels are needed to launch. Obviously, operator fusion is effective for the Attention module of fixed computing mode. More performance improvements are detailed in Table \ref{tab_modules}. Using Fused Gated Self-Attention can reduce the proportion of forward computation time for the 48-layer Evoformer from 66.46\% to 52.99\%.

Fused Gated Self-Attention can use memory read/write in-place technique in C++ operator to reduce memory peak. In fine-tuning, there will be 2 large tensor memory allocations in \emph{ExtraMSAStack}. Without Fused Gated Self-Attention, it will directly lead to OOM. Fused Gated Self-Attention adopts the memory read/write in-place technique, which reduces one of two large tensor memory allocations, and the peak memory does not exceed 40GiB. In the initial training, there is a 1288MiB memory saving.

\begin{table}[t]
\centering
\resizebox{.98\columnwidth}{!}{
\begin{tabular}{cccccc}
   \toprule
   Training Process & Operator Fusion & s/step & protein/s &  Memory  & \\
   \midrule
   \multirow{2}{*}{Initial training} & \XSolidBrush & 8.15985& 0.12255 & 24198MiB & - \\
   ~ & \Checkmark & 5.48386 & 0.18235 & 22910MiB & +48.79\% \\
   \midrule
   \multirow{2}{*}{Fine-tuning} & \XSolidBrush & - & - & OOM & - \\
   ~ & \Checkmark & 16.53586 & 0.06047 & 39246MiB & - \\
   \bottomrule
\end{tabular}
}
\caption{Performance improvements and memory savings using Fused Gated Self-Attention for initial training and fine-tuning on 1 GPUs, where batchsize is 1.}
\label{tab_fused_gating_attention_speed}
\end{table}

\subsubsection{Performance of Tensor Fusion}
To illustrate the performance improvement of tensor fusion, we conducted comparative experiments on initial training and fine-tuning respectively. As tensor fusion will involve gradient synchronization, we test with data parallelism on 8 GPUs within a single node, and the batchsize on each card is equal to 1. As shown in Table \ref{tab_fusetensor_speed}, using tensor fusion leads to a 30.71\% improvement in initial training stage. Improvement can also be seen in fine-tuning. But it is only a 10.35\% improvement due to the proportion of calculations involved tensor fusion in fine-tuning is lower. The experimental results show that tensor fusion can reduce the number of kernel launches and memory fragmentation by fusing multiple discrete tensors into a small number of large continuous tensors. Finally, we performed ablation experiments on 128 A100s, and the results show that both operator fusion and tensor fusion totaled 124.48\% performance improvements as shown in Table \ref{tab_fusion_speed}.

\begin{table}[t]
\centering
\resizebox{.98\columnwidth}{!}{
\begin{tabular}{ccccc}
   \toprule
   Operator Fusion & Tensor Fusion & s/step & protein/s  & \\
   \midrule
   \XSolidBrush & \XSolidBrush & 13.13196 & 9.74721 & - \\
   \Checkmark & \XSolidBrush & 8.92373 & 14.34378 & +47.16\% \\
   \XSolidBrush & \Checkmark & 9.20482 & 13.90576 & +42.66\% \\
   \Checkmark & \Checkmark & 5.84986 & 21.88088 & +124.48\% \\
   \bottomrule
\end{tabular}
}
\caption{Ablation study about performance improvements by fusion technical in the initial training. Tested using data parallelism on 128 A100s with total batchsize equal to 128.}
\label{tab_fusion_speed}
\end{table}

\subsubsection{Performance of Parallelism}
The author of FastFold \cite{cheng2022fastfold} has open-sourced DAP, but the open-source code only has the parallel implementation of the Evoformer module, and the related code implementations such as \emph{TemplatePairStack} and \emph{StructureModule} are not available to the public. To compare with the performance of DAP, we firstly use PaddlePaddle to reproduce the open-source code of FastFold, called PPFold. Then we add the implementation of BP on PPFold to compare the performance of DAP and BP.

Table \ref{tab_pp_speed} shows the performance comparison of FastFold and PPFold. As can be seen in the table, DAP=2 uses 2 GPUs compared to DAP=1, but there is a performance drop in both FastFold and PPFold as the last two dimensions of the inputs are relatively small with a low computational intensity. DAP splits the input into smaller tensors, and the computational intensity is not improved. In addition, a lot of additional communication overhead is introduced, resulting in a decrease in performance. BP 
splits the computing branches across different GPUs, which can be calculated in parallel while maintaining computational intensity, and only introducing a small amount of communication overhead. Thus, the performance is improved by 67.45\%.

To illustrate the effectiveness of BP, we conduct experiments with different configurations using hybrid parallelism on AlphaFold2. As shown in Table \ref{tab_hybrid_speed}, in the initial training stage, where the dimensions involved in computation are relatively small, the performance of DAP=2 drops compared to unused, showing a negative gain. When BP=2, the performance is improved by 28.87\%, and the parallel efficiency is 0.6443, showing a positive benefit. Conversely, in the fine-tuning stage, where the dimensions of MSA depth and the length of amino acid sequences increase, DAP achieves a higher performance improvement than BP by splitting larger activations across multiple GPUs for parallel computing. However, the hybrid parallelism of DAP=2 and BP=2 has higher throughput than the single parallelism of DAP=4, demonstrating that when the activation is divided to a certain size, the parallelism efficiency of BP is higher than that of DAP. Finally, we used a hybrid parallelism of BP=2, DAP=2, and DP=128 in the end-to-end fine-tuning, and achieved performance results that greatly surpassed other open-sourced methods.

\begin{table}[t]
\centering
\resizebox{.98\columnwidth}{!}{
\begin{tabular}{ccccc}
   \toprule
   Method & DAP & BP & Fwd + Bwd Time / Layer (ms) &  \\
   \midrule
   \multirow{2}{*}{FastFold} & 1 & 1 & 30.98 & - \\
   ~ & 2 & 1 & 32.25 & -3.94\% \\
   \midrule
   \multirow{3}{*}{PPFold} & 1 & 1 & 32.47 & - \\
   ~ & 2 & 1 & 33.21 & -2.22\% \\
   ~ & 1 & 2 & 19.39 & +67.45\% \\
   \bottomrule
\end{tabular}
}
\caption{FastFold VS PPFold performance comparison. Compare the total time of forward computation and backward computation for each layer. 12-layer Evoformer, data type is Float16, $head=8,B=1,N_{seq}=128,N_{res}=256,C_m=256,C_z=128$, the settings are the same as FastFold open source code. PPFold does not use asynchronous communication in DAP.}
\label{tab_pp_speed}
\end{table}

\begin{table}[t]
\centering
\resizebox{.98\columnwidth}{!}{
\begin{tabular}{ccccccc}
   \toprule
   Training Process & DAP & BP & s/step & protein/s &  Parallel Efficiency & (\%) \\
   \midrule
   \multirow{3}{*}{Initial training} & 1 & 1 & 5.54448 & 0.18036 & - & - \\
   ~ & 2 & 1 & 6.43736 & 0.15534 & 0.4306 & -13.87\% \\
   ~ & 1 & 2 & 4.30229 & 0.23243 & 0.6443 & +28.87\% \\
   \midrule
   \multirow{4}{*}{Fine-tuning} & 1 & 1 & 16.46358 & 0.06074 & - & - \\
   ~ & 2 & 1 & 11.26203 & 0.08879 & 0.7309 & +46.18\% \\
   ~ & 4 & 1 & 9.29195 & 0.10762 & 0.4429 & +77.18\% \\
   ~ & 1 & 2 & 12.82324 & 0.07798 & 0.6419 & +28.38\% \\
   ~ & 2 & 2 & 8.86957 & 0.11275 & 0.4641 & +85.62\% \\
   \bottomrule
\end{tabular}
}
\caption{Performance of end-to-end training on AlphaFold2 with hybrid parallelism within 4 GPUs, where batchsize is 1 on each card.}
\label{tab_hybrid_speed}
\end{table}

\subsubsection{Performance Analysis for More Modules}
After Operator Fusion, Tensor Fusion and hybrid parallel optimization, the performance of the AlphaFold2 model still has room for optimization, such as the \emph{StructureModule}. We decompose the AlphaFold2 model into \emph{TemplateEmbedding}, \emph{ExtraMSAStack}, \emph{EvoformerStack}, \emph{StructureModule} and other modules for performance analysis. We compare the time of each module before and after optimization, as shown in Table \ref{tab_modules}. Originally, the main computational overhead comes from \emph{EvoformerStack}, and the computational bottleneck after optimization is \emph{StructureModule}. It further explains why this paper focuses on the performance optimization of \emph{EvoformerStack}. In the follow-up work, we will focus on the performance optimization of \emph{StructureModule}, and further optimize the training and inference time of AlphaFold2.

\begin{table}[t]
\centering
\resizebox{.98\columnwidth}{!}{
\begin{tabular}{cccc}
   \toprule
   Optimization & Module & Fwd Step Time (s) & (\%)  \\
   \midrule
    \multirow{5}{*}{w/o Optimization} & TemplateEmbedding & 0.7550 & 9.84\%  \\
   ~ & ExtraMSAStack & 0.4359 & 5.68\%  \\
   ~ & EvoformerStack & 5.0971 & 66.46\% \\
   ~ & StructureModule & 1.1709 & 15.26\%  \\
   ~ & Other & 0.2096 & 2.73\%  \\

   \midrule
    \multirow{5}{*}{Fused Gated Self-Attention} & TemplateEmbedding & 0.5774 & 11.84\%  \\
   ~ & ExtraMSAStack & 0.3052 & 6.26\%  \\
   ~ & EvoformerStack & 2.5825 & 52.99\% \\
   ~ & StructureModule & 1.1482 & 23.56\%  \\
   ~ & Other & 0.2602 & 5.34\%  \\
   
   \midrule
    \multirow{5}{*}{Add BP=2} & TemplateEmbedding & 0.5839 & 14.08\%  \\
   ~ & ExtraMSAStack & 0.2678 & 6.45\%  \\
   ~ & EvoformerStack & 1.7936 & 43.25\% \\
   ~ & StructureModule & 1.1991 & 28.91\%  \\
   ~ & Other & 0.3025 & 7.29\%  \\
   
   \bottomrule
\end{tabular}
}
\caption{The forward calculation time proportion of each module before and after optimization. Since the backward calculation time in the dynamic graph is difficult to count, the table shows the forward calculation time proportion of each module of each step in the initial training. In addition, tensor fusion does not affect the calculation of each module of the AlphaFold2 model, so it is ignored. Note that \emph{ExtraMSAStack} and Evoformer use Recompute technical.}
\label{tab_modules}
\end{table}

\section{Related Work}
Since the publication of the AlphaFold2 paper \cite{jumper2021highly} and the official open-sourced inference code by JAX, multiple teams in the industry and academia have been trying to reproduce and optimize it. OpenFold \cite{Ahdritz_OpenFold_2021} reproduces almost all the features as much as possible by PyTorch. It supports training from scratch and can match the performance of the original. However, they used a larger dataset, including some 400,000 MSAs and PDB70 template hit files. Uni-Fold first reproduced trainable AlphaFold2 using JAX and recently moved the code to PyTorch. The JAX version implementation is open-sourced, but they have not released the PyTorch code. FastFold \cite{cheng2022fastfold} mentioned in their paper that they were the first to optimize AlphaFold2. In terms of distributed computing, Dynamic Axial Parallelism and Duality Async Operation aim to optimize the performance of Evoformer and save memory. In addition, they did a series of Kernel optimizations, such as \emph{Softmax} and \emph{LayerNorm}. 
However, they did not open-source end-to-end trainable code but only a parallel implementation of the Evoformer. The HelixFold proposed in this paper is also reproduced based on the inference code of AlphaFold2 on PaddlePaddle. We support training from scratch, with high throughput and low memory overhead. At the same time, we can achieve accuracy that is on par with AlphaFold2 on two datasets, TM-score is 87.7 on CASP14 and 88.8 on CAMEO.

\section{Discussion}

There are several open-sourced implementations of AlphaFold, and HelixFold is the first to offer comparable accuracy with high throughput and low memory overhead. As shown in Table \ref{tab_end2end_performance}, in end-to-end training, with the same accelerator card, HelixFold only needs 23188 GPU hours, which saves 45\% GPU hours compared to other AlphaFold2 implementations. However, to further speed up training, using more complex parallel methods, such as BP and DAP, requires more GPUs. In our implementation, we found that using a hybrid parallelism of BP-DAP-DP, requires 40063 GPU hours (72.77\% more than using DP only). In terms of economic benefits, hybrid parallelism requires more resources to obtain 1x performance improvement, and the cost performance is lower. Nevertheless, for enterprises and research institutions with sufficient computing resources, we recommend using the hybrid parallelism proposed in this paper according to their own needs.

\section{Conclusion}
Accurate and efficient protein structure prediction can benefit significantly to life science. AlphaFold2 is a cutting-edge structure prediction system demonstrating excellent accuracy in CASP14. It draws much attention from academia and industry. As the complete training of AlphaFold2 and inference of long proteins take lots of computational resources, it is a great burden for the individuals and institutions who are interested in applying AlphaFold2. Cutting down the computational cost of training and inference can enhance the development of protein-related research. For this reason, we develop an efficient protein structure prediction system. HelixFold implements AlphaFold2 with PaddlePaddle by optimizing both the computational and memory consumption. The extensive experimental results support the efficiency and effectiveness of HelixFold. We believe the publicly available HelixFold will bring more convenience to the works of life science and we will continue to further improve its efficiency and accuracy.

% Use \bibliography{yourbibfile} instead or the References section will not appear in your paper
%\clearpage
\bibliography{aaai22}

\end{document}